\documentclass[aps,prl,
            amsmath,amssymb,
            twocolumn]{revtex4}
\usepackage{bm}
\usepackage{dcolumn}
\usepackage{graphicx}
\usepackage{mathrsfs}

\newcommand{\ket}[1]{\left.\left|#1\right.\right>}

\newcommand{\eq}[1]{Eq.~(\ref{#1})}
\begin{document}

\title{Extracting molecular potentials from insufficient spectroscopic information}
\author{X.  Li\footnote{XuanLi@lbl.gov}$^1$, Sourav Dutta$^2$}
\affiliation{$^1$Chemical Sciences and Ultrafast X-Ray Science Laboratory, Lawrence Berkeley National Laboratory, Berkeley, California 94720, USA}
\affiliation{$^2$Raman Research Institute, Sadashivanagar, Bangalore 560080, India
}

\begin{abstract}
We extend our recently developed inversion method to extract excited state potentials from fluorescence line positions and line strengths. We consider a previous limitation of the method arising due to insufficient input data in cases where the relatively weaker emission data are not experimentally available.  We develop a solution to this problem by ``regenerating'' these weak transition lines via applying a model potential, e.g. a Morse potential. The result of this procedure, illustrated for the Q-branch emission from the lowest three vibrational levels of the  B($^1 \Pi)$ state of LiRb, is shown to have an error of $0.29$ cm$^{-1}$ in the classically allowed region and a global error of $5.67$  cm$^{-1}$ for $V\le E(\nu'=10)$. The robustness of this procedure is also demonstrated by considering the statistical error in the measured line intensities. 
\end{abstract}
\maketitle

\section{Introduction}
Accurate potential energy surfaces (PES) are vital to producing meaningful results in the studies of chemical reactions and molecular dynamics. So far the most reliable source of information about PES of molecules has been quantum chemical computations, of the {\it ab initio}~\cite{ab} or semi-empirical type~\cite{se1,se2}. 
For the simplest case of a diatomic molecules, it is relatively straightforward to construct the ground state potential energy curve (PEC) using experimental data available from laser induced fluorescence (LIF). In traditional LIF measurements, a molecule in an electronically excited state fluoresces to different rovibrational levels of the electronic ground state emitting light at different frequencies (determined by the energy levels) having different intensities (determined primarily by the Franck-Condon overlap). The frequency information directly reflects the vibrational and rotational energy levels of the electronic ground state and is used to construct the ground state potential using methods such as the Rydberg-Klein-Rees (RKR) method~{\cite{RKR}}. However, the information encoded in the intensity of the lines is seldom used to extract PECs. We had previously introduced a direct procedure for the
extraction ("inversion") of PEC from line strengths~\cite{JPCL2010,JCP2011, JMS2011} assuming that data for all lines, however weak in intensity, was experimentally available. Extension of this inversion formula to multi-dimensional PES has already 
been performed~\cite{JCP2011}.  Several similar methods have been recently proposed to directly reconstruct the time-dependent wave packets~\cite{Avisar,Cian2012}. In practice, however, the weak lines are often lost in the noise (we refer to a line as weak when the intensity of the line is less than 2.5\% of the strongest LIF line). Moreover, the response of the photon detectors (PMT, photodiodes etc.) may not be linear over the whole range of intensities introducing errors in the calibration of intensities. In this article, we overcome the limitation of our earlier proposal by regenerating some of the weak lines using a model potential and demonstrate that the information encoded in the intensity of the lines can be used to extract, with high accuracy, the potential energy curve for an {\it excited} electronic state.

We introduced in the previous reports~\cite{JPCL2010,JCP2011, JMS2011} the ability to reconstruct the time-independent eigenstate wave functions, $\phi_{s}(\vec{R})$, of the target excited state by expanding this in terms of eigenstate wave functions, $\chi_{i}(\vec{R})$ , of the known reference potential, e.g. the ground state. To ensure the robustness and accuracy
of this procedure, the completeness condition,
 \begin{equation}{\label{completeness}}
 \sum_{i}\left|\left < \phi_{s}(\vec{R}) \left. \right| \chi_{i}(\vec{R})\right>\right|^2\approx 1
 \end{equation}
including a summation over contributions from all ground state levels must be satisfied~\cite{JMS2011}. However,
such a requirement is not always satisfied when relatively weak transitions data, proportional to the $\left|\left < \phi_{s}(\vec{R}) \left. \right| \chi_{i}(\vec{R})\right>\right|^2$ terms in Eq.~(\ref{completeness}), are not available due to the limited signal-to-noise ratio. 
The accuracy of the extracted excited potentials becomes undermined when the completeness condition is not satisfied.  In this study, we aim at providing a solution to this limitation by ``regenerating'' these missing transition data to regain the desired accuracy of the extraction procedure. 

We choose the LiRb molecule for demonstrating our inversion method because of the high level of interest in ultracold polar LiRb molecules as borne out by several recent reports~\cite{Xpot,LiRbExp,PADutta}. Accurate potential energy curve for the electronic ground X $^1\Sigma^+$ state of the LiRb molecule is available from LIF measurements~\cite{Xpot} and, recently, the first twenty-one vibrational levels (v=0-20) of the excited B $^1\Pi$ state were studies using Fluorescence Excitation Spectroscopy (FES)~\cite{LiRbExp}. We use the full data set of the latter study~\cite{LiRbExp} to construct the RKR potential up to the v=20 level, i.e. within the classically allowed region of the first twenty-one vibrational levels (see supplementary information for the RKR potential). We then derive the potential following the current method but using LIF from only the first three vibrational levels (v=0, 1 and 2) of the  B $^1\Pi$ state. We demonstrate the success of the current method by comparing the potential we extract (using very limited experimental data) with the potential derived using the conventional RKR method (which requires a much larger data set). We find that the vibrational energy levels  calculated from the derived potential are accurate within $0.29$ cm$^{-1}$ for v=0-2. We also show that our method can be used to determine the potential in the region for which no spectroscopic information is available provided that no local perturbations are present. For example, we show that the  B $^1\Pi$ potential derived using the current method, with LIF data only from v=0-2 as input, is able to reproduce the positions of all vibrational levels up to v=10 with an accuracy of $5.7$ cm$^{-1}$. Such extensions of the potential to spectroscopically unexplored regions are very useful to experimentalists and theorists alike. For example, even for the relatively simple Li$_2$ molecule, full spectroscopic information on the 1 $^3\Sigma_g^+$ state is not available although it is a molecule of significant interest ~\cite{Semczuk2013}. In the rest of the article, we first discuss the new inversion method we proposed and the extension that we make in order to make it more accurate for practical purposes, and then establish the accuracy and usefulness of the derived potential by comparing it to the well known RKR potential.

  \begin{figure}
\includegraphics[width=2.3in, angle=-0]{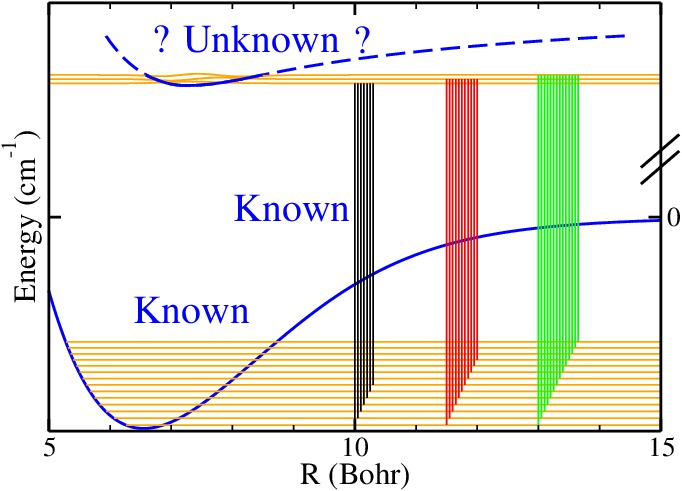}
\caption{(Color online.) Schematic plot of the aim: extract global potentials where the spectroscopy data is not available with limited fluorescence data from low-lying states only.}
 \label{fig1} \end{figure}
 
{\it Theory:} We first briefly review our inversion formula for diatomic potentials. 
In the diatomic case the potential is only dependent on the internuclear distance, denoted as
$R.$
Our objective is to extract an excited state potential $V_{\rm ex}(R)$,
assuming that we know $V_g(R),$
the ground state potential, using as input the positions and intensities
of a set of emission (or absorption) spectral lines.
A schematic illustration of the process
is depicted in Fig. \ref{fig1} for the special case of the 
LiRb B $(^1\Pi)\rightarrow$X $(^1\Sigma^+)$ emission. 

Following Ref.~\cite{JPCL2010}, when the Q branch emission data are used, we write $V_{\rm ex}(R)$
in atomic units (for which $\hbar=1$), as
 \begin{equation}
V_{\rm ex}(R)=\frac{\sum_s \sum_{i,j} d_{i,s}^* d_{j,s} \omega_{i,s}
\chi_i^*(R)\chi_j(R) }{\sum_s|\sum_i \chi_i(R)
d_{i,s}|^2} +V_g(R).
\label{DEF5}
\end{equation}
In the above, $\omega_{i,s}\equiv E_s-E_i$ are
the observed transition energies and $d_{i,s}$ are the
transition-dipole matrix elements
between $i$, the rovibrational states of the ground electronic
potential, and $s$, the rovibrational states of the excited state
potential. The transition dipole matrix elements are
defined as
\begin{equation}{\label{DIPOLE1}}
d_{i,s}\equiv \int dR \chi^*_i(R) \mu_{{\rm e},g}(R)
\phi_s (R)\approx
\overline{\mu}_{{\rm e},g}f_{is}
\end{equation}
where $f_{is}\equiv \int dR \chi^*_i(R)\phi_s (R).$
The last equality spells out the Franck Condon Approximation
(FCA)\cite{FCA}.

\begin{figure}
\includegraphics[width=2.3in, angle=-0]{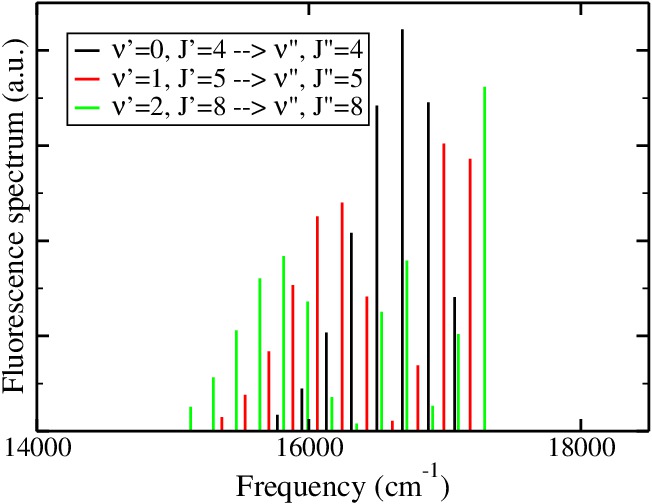}
\caption{(Color online).  Sample emission spectrum (Q-branch only) from the $\ket{\nu'=0, J'=4}$, $\ket{\nu'=1, J'=5}$, and $\ket{\nu'=2, J'=8}$.}
 \label{fig2} \end{figure}

 
 {\it Results and discussion:} In this study, we envision an experiment similar to Ref.~\cite{LiRbExp}. Our particular objective
in this study is to use the fluorescence data from $\nu'=[0,1,2]$ to extract the potential 
in the spectra range of $E(\nu'=2)\le V_B(R)\le E(\nu'=20)$. In order to avoid systematic error and demonstrate the capability of the proposed method, we numerically generate the fluorescent data based on the Frank-Condon
overlaps between the wave functions of the B-state~\cite{Bpot} and the X-state~\cite{Xpot} potentials. The numerically generated emission spectrum is shown in Fig.~\ref{fig2}, which contains both the $|d_{i,s}|^2$ line intensities and $\omega_{i,s}$ line positions 
of the B($^1\Pi$)$\rightarrow$X($^1\Sigma^+$) band (this may be compared with the experimentally measured intensities reported in Fig. 2 of Ref.~\cite{LiRbExp}) .
 Here, only $7$, $11$, and $14$ lines are used and plotted for $\nu'=0$, $1$, and $2$, respectively, which are greater than $2.5\%$ of the maximum signal to simulate the experimental condition.
 
\begin{figure}
\includegraphics[width=2.3in, angle=-0]{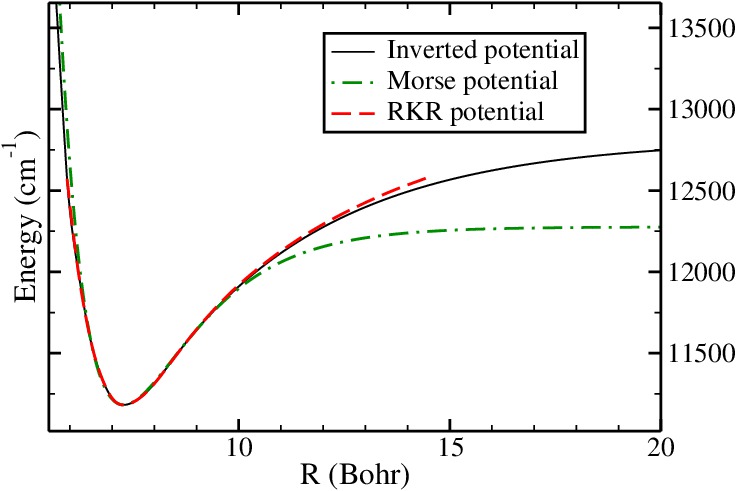}
\includegraphics[width=2.3in, angle=-0]{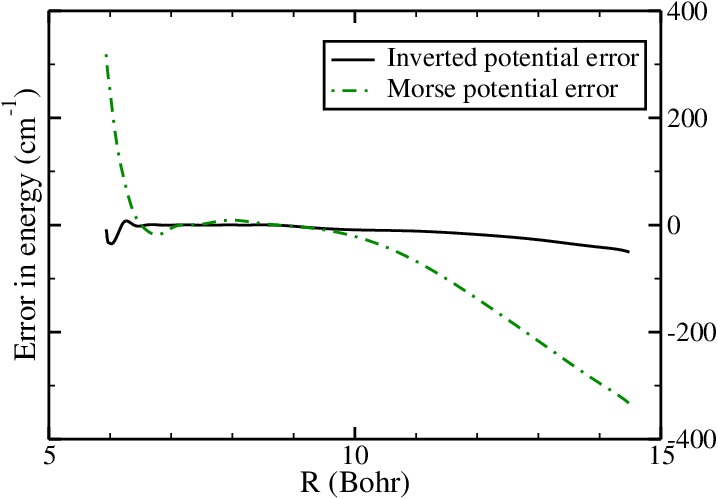}
\caption{(Color online). (a) Comparison of the inverted potential using  \eq{DEF5} with the fitted Morse potential in the interested region; the red dashed line is the RKR potential. (b) Internuclear distance dependent error of the inverted potential (black solid) and the fitted Morse potential (green dot-dash). }
 \label{fig3} \end{figure}

 One way to extract the potential beyond the classical turning points of the $\nu'=2$ state is to model the potential as an Morse potential with parameters chosen to match the energies of the lowest three vibrational  states. 
 Three such parameters for the Morse potential  
 \begin{equation}
V_M(R)=T_e+D_e(e^{-2\beta (R-R_e)}-2e^{-\beta (R-R_e)}).
\end{equation} 
 are $T_e$, $D_e$ and $\beta$, where $T_e$ is the asymptote of the potential, $D_e$ is the dissociation energy and $\beta$ is a parameter dependent on the force constant.
Note, the $R_e$ parameter is varied to match the line intensity profile, $|d_{i,s}|^2$ in Fig.~\ref{fig2}. Such a Morse potential gives the correct description of the potential for $V(R)\le E(\nu'=2)$, shown as the green dot-dash line in Fig.~\ref{fig3}; but it starts to deviate significantly from the RKR potential at higher energies, shown as the red dashed line in Fig.~\ref{fig3}. Such behavior is expected because the Morse potential is derived based on the line information of only the lowest three states.

An alternative way to greatly extend the extracted potential beyond the classical turning points is to use the line intensities. The line intensities depend on the FC overlap between the ground and excited state wave functions. Since the wave function of a vibrational level can extend well beyond turning points into the classically forbidden region for that particular vibrational level, the FC overlap and hence the line intensity contain information regarding this classically forbidden region.  
However, such useful information is reflected by the weak transitions which are usually hard to distinguish from the background noise. Hence, we propose a method to compensate for these missing data by exploiting the Morse potential deduced earlier.  Once the Morse potential is deduced from the line positions, 
we can compute
the energy eigenfunctions, i.e. the rovibrational wave functions 
$\phi_s(R)$ corresponding to the excited state potential
$V_{ex}(R)$, by solving the time-independent Schr\"odinger equation. 
These rovibrational wave functions
can then be used to compute the Frank-Condon overlap and the relevant sign information~\cite{JPCL2010}. These overlaps are used to ``make up'' for the missing/weak lines; the combination of 
the overlaps computed from using the Morse potential and those derived from the measured strong emission lines
are used in \eq{DEF5} to yield the final form of the potential. Once the extracted potential is generated, we extrapolate the potential beyond a point where the density of the extracted wave function, $\sum_i \chi_i(R)
d_{i,s}$ in \eq{DEF5},  becomes small.
 
  We use this method to 
extract the B($^1\Pi$) potential. The RMS errors of the inverted potentials, using the fluorescence data from $\nu'=0,~1, ~2$,
are given in Table ~\ref{POTRMS} for the $V(R)<E(\nu'=2)$,
$V(R)<E(\nu'=5)$, $V(R)<E(\nu'=10)$ and $V(R)<E(\nu'=20)$ regions. 
 The final form of the extracted potential is plotted and shown as the black solid line in Fig.~\ref{fig3}a; its R-dependent error compared with the RKR potential is shown in Fig.~\ref{fig3}b. It is obvious
 that the extracted potential gives a very accurate description of the classical allowed region, $V(R)\le E(\nu'=2)$;
 in addition, this potential is fairly accurate in the classical forbidden region.

 \begin{table}[h]
  \caption{Performance of the inverted potential in various regions based
  on the fluorescence data from $\nu'=0,~1,~2$.}
 \begin{tabular}{|c|c|}
    \hline
    Region & RMS error\\
    \hline\hline
  V$<$E$(\nu'=~2)$ & $~0.29$cm$^{-1}$ \\ \hline
    V$<$E$(\nu'=~5)$ & $~1.62$cm$^{-1}$ \\ \hline
   V$<$E$(\nu'=10)$ & $~5.67$cm$^{-1}$ \\ \hline
    V$<$E$(\nu'=20)$ & $25.08$cm$^{-1}$ \\ \hline 
        \hline
    \end{tabular}
    \label{POTRMS}
\end{table}

\begin{figure}
\includegraphics[width=2.3in, angle=-0]{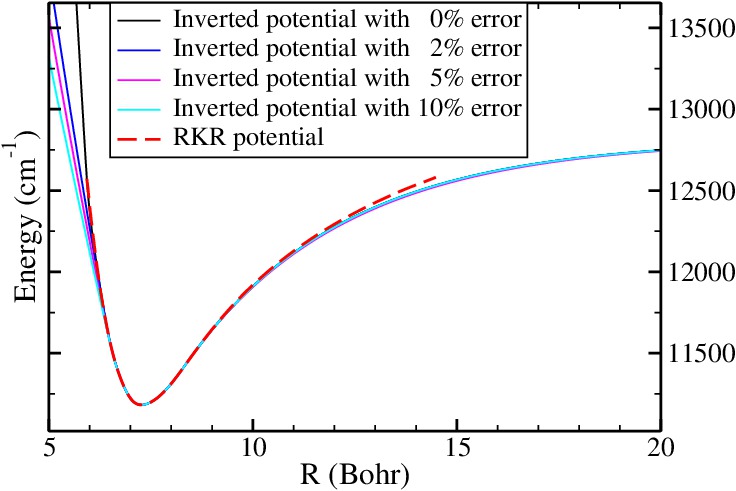}
\includegraphics[width=2.3in, angle=-0]{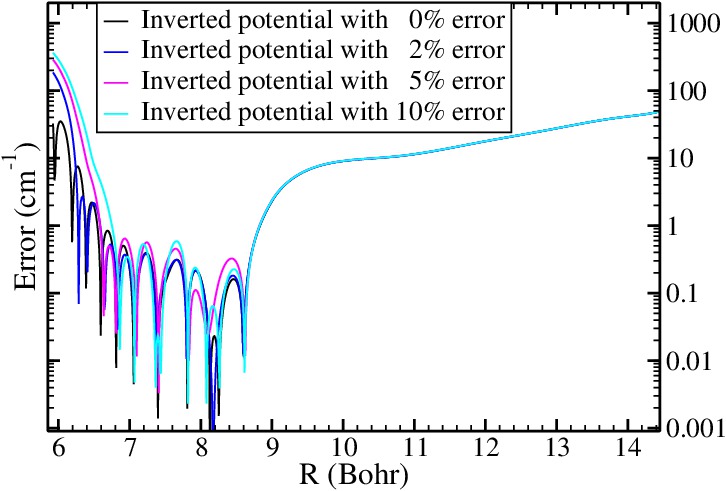}
\caption{(Color online). (a) Comparison of the inverted potential based on fluorescence data with random errors of, $0\%$, $2\%$, $5\%$ and $10\%$. (b) the R-dependent error of the inverted potential when compared with the RKR potential.}
 \label{fig4} \end{figure}

A question of major importance is the robustness of this procedure against common experimental sources of errors such as errors in the line intensities, which are directly related to the $|d_{i,s}|^2$ terms in \eq{DEF5}. 
In this study, we focus on the statistical random errors and, to address this issue,e we introduced random errors to the $30$ $|d_{i,s}|^2$ data in Fig.~\ref{fig2} and repeated the extraction of the potentials in the presence of such errors. The errors in $|d_{i,s}|^2$ were generated in a random fashion of  $2,~5,$ and $10\%$ rms errors relative to the average $|d_{i,s}|^2$ values. Note, the line intensities for the missing/weak $63$ lines deduced from the Morse potential are not randomized since they are based on the line positions, which are, in general, of high accuracy. 
In Fig.~\ref{fig4}, we show the average B($^1\Pi$) numerical potential and its deviation from the RKR potential 
at different R values extracted from such data for the above magnitudes of errors.  Fig.~\ref{fig4} demonstrates a remarkable robustness against inaccuracies in the experimental fluorescence data. The fluctuation in the magnitude of the errors in Fig.~\ref{fig4} as a function of the internuclear distance have been observed previously~\cite{JPCL2010}. The origin of such an effect is still unknown: it has no obvious correlation with the nodal structure of the wave function. The RMS errors of the average B($^1\Pi$) numerical potential, using the fluorescence data from $\nu'=0,~1, ~2$,
are given in Table ~\ref{POTRMS2} for the $V(R)<E(\nu'=2)$,
$V(R)<E(\nu'=5)$, $V(R)<E(\nu'=10)$ and $V(R)<E(\nu'=20)$ regions.

\begin{table}[h]
  \caption{Performance of the inverted potential in various regions based
  on the fluorescence data from $\nu'=0,~1,~2$ with different random errors in the relative intensities of the fluorescence data: (a) $2\%$, (b) $5\%$ and (c) $10\%$.}
 \begin{tabular}{|c|c|c|c|}
    \hline
    Region & RMS error  & RMS error & RMS error  \\
     & for data set 1$^a$  & for data set 2$^b$ & for data set 3$^c$  \\
    \hline\hline
    V$<$E$(\nu'=~2)$ & $~0.23$cm$^{-1}$ & $~~0.32$cm$^{-1}$ & $~0.78$cm$^{-1}$\\ \hline
    V$<$E$(\nu'=~5)$ & $~1.62$cm$^{-1}$& $~~2.56$cm$^{-1}$ & $~5.56$cm$^{-1}$ \\ \hline
    V$<$E$(\nu'=10)$ & $~8.42$cm$^{-1}$& $~21.30$cm$^{-1}$ & $39.54$cm$^{-1}$ \\ \hline
    V$<$E$(\nu'=20)$ & $40.88$cm$^{-1}$& $71.42$cm$^{-1}$ & $107.59$cm$^{-1}$ \\ \hline 
        \hline
    \end{tabular}
        \label{POTRMS2}
\end{table}

To summarize, we extend the recently developed inversion method for the accurate extraction
of excited state potentials from fluorescence line positions and line strengths. 
We consider one of the limitation to our previous work arising when the relatively weaker
emission data are not available. We develop a method to address this issue 
by ``regenerating'' these weak transition lines by first  generating a model potential,
e.g. a Morse potential based on the lowest few vibrational energy lines. We then numerically compute
the first-order wave function to yield the Frank-Condon overlaps with those of the known
ground state potential. This procedure, illustrated for the LiRb B($^1 \Pi)$$\rightarrow$X($^1\Sigma^+$)
Q-branch emission from the lowest three vibrational levels, result in a $0.29$ cm$^{-1}$ error in the classically allowed region and a global error of $5.67$  cm$^{-1}$ for $V\le E(\nu'=10)$. We also demonstrate the robustness of this procedure by considering the statistical errors in the measured line intensities.  We consider this method as an effective method to deduce accurate spectroscopic information in the previously unknown spectral region.

We are grateful for the discussion with Moshe Shapiro and Cian Menzel-Jones.


\begin{thebibliography}{50}
\bibitem{ab}{See, e.g., I. N. Levine, {\it Quantum Chemistry}, Prentice Hall, Englewood Cliffs, NJ, 1991 and references therein.}
\bibitem{se1}{F. O. Ellison, J. Am. Chem. Soc. {\bf 85}, 3540 (1963); J. C. Tully, J. Chem. Phys. {\bf 58}, 1396 (1973); E. Steiner, P. R. Certain, P. J. Kuntz, J. Chem. Phys. {\bf 59}, 47 (1973); C. W. Eaker, C. A. Parr, J. Chem. Phys. {\bf 64}, 1322 (1976); B. Kendrick, R. T. Pack, J. Chem. Phys. {\bf 102}, 1994 (1995); X. Li, D. A. Brue, G. A. Parker, J. Chem. Phys. {\bf 129}, 124305 (2008).}
\bibitem{se2}{P. J. Kuntz, J. Valldorf, Z. Phys. D {\bf 8}, 195 (1988).}
\bibitem{RKR}{R. Rydberg, Z. Phys. {\bf 73}, 376 (1931); R. Rydberg, Z. Phys. {\bf 80} 514 (1933); O. Klein, Z. Phys. {\bf 76}, 226 (1932); A. L. G.
Rees, Proc. Phys. Soc. {\bf 59}, 998 (1947).}
\bibitem{JPCL2010}{X. Li, C. Menzel-Jones, and M. Shapiro, J. Phys. Chem. Lett. {\bf 1} 3172 (2010); X. Li, C. Menzel-Jones, D. Avisar, M. Shapiro Phys. Chem. Chem. Phys. {\bf 12} 15760 (2010); X, Li, M. Shapiro
Israel J. Chem. {\bf 52} (5), 407 (2012).}
\bibitem{JCP2011}{X. Li and M. Shapiro, J. Chem. Phys. {\bf 134}, 094113 (2011). }
\bibitem{JMS2011}{C. Menzel-Jones, X. Li, M. Shapiro, J. Mole. Spectro. {\bf 268} 221 (2011).}
\bibitem{Avisar}{D. Avisar, D. J. Tannor, Phys. Rev. Lett., {\bf 106}, 170405 (2011);D. Avisar, D. J. Tannor, Faraday Discuss.,{\bf 153}, 131 (2011); D. Avisar, D. J. Tannor, J. Chem. Phys. {\bf 136}, 214107 (2012).}
\bibitem{Cian2012}{C. Menzel-Jones, and M. Shapiro, J. Phys. Chem. Lett., {\bf 3} 3353 (2012); C. Menzel-Jones, and M. Shapiro, J. Phys. Chem. Lett., {\bf 4} 3083 (2013).}
\bibitem{Xpot}{M. Ivanova, A. Stein, A. Pashov, H. Kn\"{o}ckel, and E. Tiemann, J. Chem. Phys. {\bf 134}, 024321 (2011)}
\bibitem{LiRbExp}{S. Dutta, A. Altaf, D.S. Elliott, 
Y. P. Chen, Chem. Phys. Lett. {\bf 511} 7, (2011).}
\bibitem{PADutta}{ S. Dutta, D. S. Elliott, Y.P. Chen, Euro. Phys. Lett. {\bf 104} 63001 (2013); S. Dutta, J. Lorenz, A. Altaf, D. S. Elliott, Y.P. Chen, Phys. Rev. A, {\bf 89} 020702 (2014).}
\bibitem{Semczuk2013}{M. Semczuk, {\it et al.}, Phys. Rev. A {\bf 87}, 052505 (2013). }
\bibitem{FCA}{See, e.g., G. Herzberg, {\it Molecular spectra and molecular
structure, Vol. I. Spectra of diatomic molecules} (Van Nostrand Reinhold,
New York, 1950)}
\bibitem{Bpot}{The RKR potential is generated based on the experiment done in Ref.~\cite{LiRbExp} and also provided in the supplemental material.}
\end{thebibliography}
\end{document}